# Comment on "Dynamics in an oil-continuous droplet microemulsion as seen by quasielastic scattering techniques" by T. Hellweg et al.

V. Lisy

Department of Biophysics, P.J. Safarik University, Jesenna 5, 041 54 Kosice, Slovakia

___

It is shown that the interpretation of quasielastic scattering experiments on droplet microemulsions in the paper by T. Hellweg et al., *Phys. Chem. Chem. Phys.*, 2000, **2**, 5168, contains serious shortcomings and should be revised.

___

In the recent paper[1] the direct determination of the elasticity coefficients $\kappa$ and $\bar{\kappa}$ of the surfactant film for an oil-continuous droplet microemulsion has been presented. These constants have been attempted using three experimental techniques. The small-angle neutron scattering (SANS) was used to get the polydispersity of the droplets in radii, $p^2$, and the mean droplet radius in the sample, $R_m$. The dynamic light scattering (DLS) measurements were used to obtain the self-diffusion coefficient of the droplets, $D_0$, and also the polydispersity. From the neutron spin-echo experiment (NSE) the relaxation time $\tau_2$ of the lowest mode of the droplet fluctuations in the shape has been accessed. By a combined analysis of the found quantities both $\kappa$ and $\bar{\kappa}$ have been determined. From the methodical point of view this work represents a perspective way of using alternative techniques to gain information about the parameters of the surfactant film. However, the realization of this program contains shortcomings that make the interpretation of the experiments incorrect so that the extracted values of $\kappa$ and $\bar{\kappa}$ are not reliable. To prove this statement we begin with the intermediate scattering function (eqn.(16) in Ref.[1]) $I(q,t)$ that describes the quasielastic scattering of neutrons on a spherical droplet fluctuating in the shape. Equation (16) corresponds to the perfect shell contrast (when the scattering length densities $\rho_1$ and $\rho_2$ of the bulk fluids are equal) and to the lowest nonvanishing approximation in the small thickness of the surface layer $d$. $I(q,t)$ consists of two parts. The time-independent part is determined by the function

$$f_0(x) = j_0^2(x) + j_0(x)[(4-x)j_0(x) - 2j_1(x)]\sum_{l>1}(2l+1)\langle u_{l0}^2 \rangle / 4\pi, \quad x = qR,$$

where $u_{l0}$ are fluctuation amplitudes of the radius.[1] The first term here describes the static scattering from a nonfluctuating spherical shell of the radius $R$. The second term accounts for the shape fluctuations. This contribution should be calculated up to the second order in the small quantities $u_{l0}$. This has been done in our paper[2] for arbitrary scattering length densities and $d$. The correct result for $f_0(x)$ in the limit $d\to 0$ and for $\rho_1=\rho_2$ differs from the above given expression by the factor before the sum which should be $-j_0^2(x)(2+x^2)$. In Ref.[2] all the second-order contributions from the fluctuations have been summed in $I(q,t)$. This means that $I(q,t)$ contains also a term $\sim \langle u_{00} \rangle$ arising due to the conservation of the droplet volume (assumed also in Ref.[1]). In Ref.[3] our calculations have been confirmed except that the contribution due to the volume constraint was missed. Even in this case the result for $I(q,t)$ differs from that in Ref.[1] since, in the mentioned limit, the factor before the sum in $f_0$ is now $-j_0(x) \times [2xj_1(x)+(x^2-2)j_0(x)]$.[3] Notice that if the droplet volume is conserved, the terms $\sim \langle u_{00} \rangle$ cannot be neglected when any physical quantity is calculated to the second order in the fluctuation amplitudes. The authors[1] state (without any evidence) that the polydispersity is erroneously counted twice in our approach[2]. To make clear with this question we only remark that $\langle u_{00} \rangle$ is not the polydispersity that is often treated as $\langle u_{00}^2 \rangle$.[1] It simply follows from the condition for the droplet volume, $V = 4\pi R^3/3$, hence $u_{00} = -(4\pi)^{-1/2} \sum_{l>1,m} |u_{lm}|^2$.[4] Omitting the $\sim u_{00}$ term the volume would be $V = (4\pi R^3/3)\left[1+(3/4\pi)\sum_{l>1,m}|u_{lm}|^2\right]$. One can assume a different meaning of the radius $R$ (not the equivalent-volume radius); then $R$ will implicitly depend on the fluctuation amplitudes that should be reflected in eqns.(16)-(18).[1] In our approach the polydispersity appears only after the averaging over the distribution of the droplets in radii that follows from the microemulsion thermodynamics. For example, the averaged droplet volume is $\langle V \rangle_R \approx (4\pi/3)R_m^3(1+3p^2)$. The use of a different distribution, like the Schulz one[1], makes the calculations inconsistent with the theory of the droplet formation.

Note that expressing $I(q,t)/I(q,0)=\exp(-D_0 q^2 t)[a+(1-a)\exp(-t/\tau_2)]$ and adjusting the parameters $a$ and $\tau_2$ to the NSE data, a quantitative disagreement with eqn.(16) has been observed already in the paper[5] by one of the authors.[1] The coefficient $D_0$ was determined from DLS and from $\tau_2$ the sum $4\kappa \cdot \bar{\kappa}$ was computed according to eqn.(19).[1] However, the used expression for $\tau_2$ is not correct: it is not applicable to incompressible shells assumed in[1], as it was already pointed out in Refs.[6,7] (see also[8]). Moreover, the account for the higher ($l>2$) modes is essential in the calculations of the scattering functions.[2] The use of eqn.(20) is thus not justified: it brings an uncontrolled error in the determination of $\kappa$ and $\bar{\kappa}$. As to the self-diffusion coefficient $D_0$ determined from DLS, it was shown that its correct extraction from the experiments requires a description of the scattering that takes into account different refractive indices of the oil, water, and surfactant.[2] This was not done in Ref.[1]. The description of the quasielastic experiments is thus flawed. The interpretation of SANS is not correct as well. Treating SANS within the model of a diffuse scattering length boundary (note the difference with the model used for the description of NSE), the droplet fluctuations are fully ignored. In the description of SANS within the model of droplets with sharp boundaries it would mean that the formfactor of the static scattering, $P(q)=I(q,0)$, is determined simply by $f_0(x)=j_0^2(x)$. The influence of the fluctuations is however essential as analyzed in detail previously.[2] The measured SANS signal is sensitive not only to the layer thickness, the polydispersity, and the mean radius; it is also significantly affected by the value of $\kappa$. The analysis in Ref.[1] implicitly assumes $\kappa$ to be so large that the fluctuations can be neglected. This contradicts to the treatment of the inelastic scattering experiments. It also makes the determination of the combination $2\kappa+\bar{\kappa}$ from $p^2$ (obtained when eqns.(12) and (13) with no fluctuation contributions are fitted to the SANS data)[1] doubtful.

In conclusion, we have demonstrated that while the method for the determination of the basic microemulsion parameters presented in Ref.[1] is perspective, the interpretation of all the experiments carried out in the discussed work is flawed and should be revised in a number of points.